# Evidences of extragalactic origin and planet engulfment in the metal-poor twin pair HD 134439/HD 134440

Henrique Reggiani[1,2★] and Jorge Meléndez[1★]

[1]*Universidade de São Paulo, Instituto de Astronomia, Geofísica e Ciências Atmosféricas, IAG, Departamento de Astronomia, Rua do Matão 1226, Cidade Universitária, 05508-900, SP, Brazil*
[2]*Max-Planck Institute for Astronomy, Konigstuhl 17, D-69117 Heidelberg, Germany*

Accepted 2018 January 8. Received 2018 January 5; in original form 2017 October 25

## ABSTRACT

Recent studies of chemical abundances in metal-poor halo stars show the existence of different populations, which is important for studies of Galaxy formation and evolution. Here, we revisit the twin pair of chemically anomalous stars HD 134439 and HD 134440, using high resolution ($R \sim 72\,000$) and high S/N ratio (S/N $\sim$ 250) HDS/Subaru spectra. We compare them to the well-studied halo star HD 103095, using the line-by-line differential technique to estimate precise stellar parameters and LTE chemical abundances. We present the abundances of C, O, Na, Mg, Si, Ca, Sc, Ti, V, Cr, Mn, Co, Ni, Cu, Zn, Sr, Y, Ba, La, Ce, Nd, and Sm. We compare our results to the precise abundance patterns of Nissen & Schuster (2010) and data from dwarf Spheroidal galaxies (dSphs). We show that the abundance pattern of these stars appears to be closely linked to that of dSphs with [$\alpha$/Fe] knee below [Fe/H] $< -1.5$. We also find a systematic difference of $0.06 \pm 0.01$ dex between the abundances of these twin binary stars, which could be explained by the engulfment of a planet, thus suggesting that planet formation is possible at low metallicities ([Fe/H] $= -1.4$).

## 1 INTRODUCTION

Uncovering the details of our Galaxy's formation is one of the key studies in modern astronomy. Nissen & Schuster (2010, hereafter NS10) identified the existence of two populations in [$\alpha$/Fe], [Na/Fe], [Ni/Fe], and in stellar kinematics, around the solar neighbourhood for metallicities in the range of $-1.6 \leq$ [Fe/H] $\leq -0.8$, and attributed the differences between the populations to the environments in which these stars formed. While one usually attributes the high-$\alpha$ population as stars formed locally in the Milky Way, the low-$\alpha$ stars could have been formed in dwarf galaxies and latter accreted by our larger Galaxy.

In subsequent studies, Nissen & Schuster (2011, hereafter NS11) showed that the distinction between the two populations also extends to other chemical elements ([Cu/Fe], [Zn/Fe], and [Ba/Y]), and Ramírez, Meléndez & ChanameÌ (2012) and Fishlock et al. (2017) extended the previous studies to O, Sc, and neutron-capture elements. NS11 attributed this separation to the different star formation rates (SFRs) of the two environments. In a high SFR environment, mostly massive stars and type II supernovae would contribute to the chemical enrichment and the ratio [$\alpha$/Fe] increases to form the high-$\alpha$ population *in situ*, and in a low-SFR environment, such as that of dwarf galaxies, a slower chemical enrichment with type Ia supernovae and low-mass AGB stars results in lower [$\alpha$/Fe] ratios and other distinct abundance ratios.

The pair of high proper motion stars HD 134439 and HD 134440 were suggested to be accreted from another galaxy by Carney et al. (1996), based on an analysis of their kinematics. An abundance study by King (1997) showed low [$\alpha$/Fe] ratios, reinforcing the idea that the stars were not formed in the Milky Way, but rather in an environment like a dSph. Shigeyama & Tsujimoto (2003) proposed that stars with low [$\alpha$/Fe] ratios may have engulfed planets or planetesimals. Analysis of more elements by Chen & Zhao (2006) confirmed a distinctive low abundance of $\alpha$-elements, concluding that the environment in which these stars were formed had few type II SNe and a high dust-to-gas ratio, meaning that the pair may have been formed in a dusty dSph medium. Based on the analysis of Be, C, N, O, Ag, and Eu, complemented with previous abundances of other elements, Chen, King & Boesgaard (2014) discussed the above scenarios (dSph and planet accretion), and proposed that although an environment such as a dSph could be responsible for some of the abundance ratios, some undetermined nucleosynthetic processes may be responsible for some apparent anomalous abundance ratios in the neutron-capture elements.

The puzzling abundance ratios that Chen & Zhao (2006) found for these stars could be connected to the recent findings of different halo

★ E-mail: hreggiani@gmail.com/hreggiani@usp.br (HR); jorge.melendez@iag.usp.br (JM)

populations (NS10) or the different patterns seen in stars of dSphs (e.g. Tolstoy, Hill & Toii 2009; Suda et al. 2017). Indeed, as pointed out by Suda et al. (2017), the distinctive $\alpha$-elements abundance pattern ([$\alpha$/Fe] $< 0$) is also observed in stars of dSphs such as Fornax (Letarte et al. 2010; Lemasle et al. 2014), Carina (Lemasle et al. 2012; Venn et al. 2012; Fabrizio et al. 2015), Sculptor (Geisler et al. 2005), Draco (Shetrone, Bolte & Stetson 1998; Shetrone, Côté & Sargent 2001; Cohen & Huang 2009), Sextans (Shetrone et al. 2001; Aoki et al. 2009), and Sagittarius (Bonifacio et al. 2000, 2004; Sbordone et al. 2007) (see fig. 10 of Suda et al. 2017).

In light of recent measurements of chemical abundances in the halo and dSphs, we revisit the high proper-motion pair HD 134439 and HD 134440, using new data and precise abundances obtained through the differential technique. In Section 2, we detail the data used in this work and the spectra reduction processes. In Section 3, we describe the determination of stellar parameters. In Section 4.1, we compare the stellar chemical composition to the results of the low- and high-$\alpha$ halo stars of NS10, NS11, Ramírez et al. (2012), and Fishlock et al. (2017). In Section 4.2, we discuss the connection with dSphs using abundances from Shetrone et al. (2003), Geisler et al. (2005), Monaco et al. (2005), and Letarte et al. (2010). In Section 4.3, we assess possible trends with condensation temperature. In Section 4.4, we discuss a possible planet signature and our conclusions are presented in Section 5.

## 2 OBSERVATIONS AND DATA REDUCTION

We observed three stars (HD 103095, HD 134439, and HD 134439) at the 8.2 m SUBARU telescope, with the high dispersion spectrograph HDS (Noguchi et al. 2002), Subaru programme (9S16A-TE005/o162060), under the Gemini time exchange programme (GU2016A-005). We observed the targets on 05/27/2016, with the standard HDS set-ups Rb and Yc, with a total wavelength coverage of 4400 Å–7950 Å. The slit width was set to 0.5 arcsec, corresponding to a resolving power of $R = 72\,000$. The exposure times were 100, 1000, and 1500 s for HD 103095, HD 134439, and HD 134440, respectively, yielding an S/N $\sim$ 250 at 5000 Å.

In order to make a good differential analysis, it is necessary to use a very well-known star as a standard point of comparison for the measurements. With that in mind, we observed the star HD 103095 as our standard for the objects of our study (HD 134439/134440 pair). This star has been measured several times and there are detailed spectroscopic analysis that can be traced back to as far as Tomkin (1972, and references therein). The stellar parameters of the chosen star must be as similar as possible to the other objects of study, thus rendering HD 103095 a very good choice for this study (as will be further shown later in the paper).

We also obtained a spectrum from the ESO archive for the star HD 163810 (ESO programme 071.B-0529). This UVES/VLT data have a wavelength coverage of 4000–7800 Å, with a spectral resolution of $R = 47\,000$ and S/N $\sim$ 250 at 5000 Å.

HD 163810 was analysed in NS10, NS11, and Ramírez et al. (2012). Their measurements place the pattern of this star in the low-$\alpha$ population. Their adopted stellar parameters ($T_{\rm eff} = 5501$ K, log $g = 4.56$ dex, [Fe/H] $= -1.20$ dex, and $v_T = 1.30$ km s$^{-1}$) are relatively close to the stellar parameters of the binary pair we are studying, so we also analyse this star to use its abundances as a consistency check of our results.

The carbon measurements of HD 103095, HD 134439, and HD 134440 were done via spectral synthesis of the 4300 Å carbon *G*-band feature, using HIRES/Keck data from the KOA archive,[1] project N31H. To synthesize the CH lines, we used the spectral region 4295–4350 Å, which has an S/N $\sim$ 100.

The HDS/Subaru data were reduced in real-time by the Subaru staff using IRAF scripts, performing flat, bias corrections, spectral order extractions, and wavelength calibrations. Later, we also corrected the spectra for barycentric and radial velocities and normalized the spectra. The UVES/VLT data obtained had been reduced by the ESO pipeline, which performed order extraction, flat-field and bias corrections, and wavelength calibration. The remaining processing (barycentric and radial velocity corrections) were done using the IRAF package for PYTHON, PYRAF,[2] and the normalization of the spectrum was done with IRAF. The HIRES/Keck data are provided in reduced format already, processed through the MAKEE[3] package. Further processing (Doppler correction and normalization) was done with IRAF.

## 3 STELLAR PARAMETERS AND CHEMICAL ABUNDANCES

We have performed manual EW measurements via the *splot* task in IRAF, using Gaussian profile fitting. It is also important to stress that a given line was measured consecutively in all three stars, setting a consistent continuum, resulting thus in reliable precise measurements. An example of the linelist with the measured EW can be seen in Table A1, and the full linelist can be found online. The EWs were analysed employing the semi-automatic code q2 (Ramírez et al. 2014), which uses EW to fit abundances through curves-of-growth employing the ABFIND function of MOOG (Sneden 1973), with MARCS plane-parallel 1D model atmospheres (Gustafsson et al. 2008) to estimate stellar parameters and abundances.

The determination of stellar parameters and chemical abundances was done via a line-by-line differential analysis (Meléndez et al. 2012; Yong et al. 2013; Ramírez et al. 2015; Reggiani et al. 2016). As mentioned, we chose to observe HD 103095 as the standard star of the differential analysis. We adopted the stellar parameters of HD 103095 based on the recent detailed work by Sitnova et al. (2015): $T_{\rm eff} = 5100 \pm 65$ K, log $g = 4.65 \pm 0.08$ dex, $v_T = 0.90 \pm 0.05$ km s$^{-1}$, and we adopted [Fe/H] $= -1.35 \pm 0.08$ dex from our Fe I and Fe II line measurements, which agrees within the errors with the [Fe/H] $= -1.26 \pm 0.08$ iron abundance determined by Sitnova et al. (2015).

Using the Fe I and Fe II abundances as reference, we employed the differential technique and obtained the model atmospheric parameters of the other three stars; these are summarized in Table 1. The stellar parameters we determined for star HD 163810 ($T_{\rm eff} = 5526$ K, log $g = 4.56$ dex, [Fe/H] $= -1.26$ dex, and $v_T = 0.99$ km s$^{-1}$) are in good agreement with the atmospheric parameters derived by NS10 (shown in Section 2). This excellent agreement between the stellar parameters of HD 163810 from NS10 and ours provides reliability to the comparison of our abundance pattern to theirs.

The differential chemical abundances of 24 elements, including some neutron capture elements (C, O, Na, Mg, Si, K, Ca, Sc, Ti, V,

[1] The Keck Observatory archive (KOA) is a joint development between the W. M. Keck Observatory (WMKO) located in Waimea, Hawaii and the NASA Exoplanet Science Institute (NExScl) located in Pasadena, California. https://www2.keck.hawaii.edu/koa/public/koa.php

[2] PYRAF is a product of the Space Telescope Science Institute, which is operated by AURA for NASA.

[3] http://www.astro.caltech.edu/~tb/makee/

**Table 1.** Stellar parameters and chemical abundances of the sample. The standard star is highlighted. $T_{\rm eff}$ is in K, log *g* in dex, [Fe/H] in dex, $v_T$ in km s$^{-1}$, and the [X/Fe] in dex.

| [X/Fe] | **HD 103095** | HD 163810 | HD 134439 | HD 134440 |
|---|---|---|---|---|
| $T_{\rm eff}$ | 5100 | 5526 | 5084 | 4946 |
| $\sigma T_{\rm eff}$ | 65 | 47 | 27 | 26 |
| log *g* | 4.65 | 4.56 | 4.66 | 4.68 |
| $\sigma$log *g* | 0.08 | 0.09 | 0.06 | 0.07 |
| [Fe/H] | −1.35 | −1.26 | −1.43 | −1.39 |
| $\sigma$[Fe/H] | 0.08 | 0.03 | 0.02 | 0.02 |
| $v_T$ | 0.90 | 0.99 | 1.22 | 1.17 |
| $\sigma v_T$ | 0.05 | 0.15 | 0.08 | 0.06 |
| C | −0.35 | – | −0.15 | −0.13 |
| $\sigma$C | 0.100 | – | 0.150 | 0.130 |
| O | 0.549 | 0.691 | ≤0.040 | ≤−0.020 |
| $\sigma$O | 0.077 | 0.051 | 0.039 | 0.051 |
| Na | −0.262 | −0.331 | −0.382 | −0.314 |
| $\sigma$Na | 0.050 | 0.096 | 0.039 | 0.031 |
| Mg | 0.117 | 0.087 | −0.026 | −0.067 |
| $\sigma$Mg | 0.044 | 0.032 | 0.022 | 0.023 |
| Si | 0.089 | 0.106 | −0.003 | 0.027 |
| $\sigma$Si | 0.071 | 0.045 | 0.048 | 0.049 |
| K | 0.339 | 0.454 | 0.143 | 0.136 |
| $\sigma$K | 0.094 | 0.129 | 0.059 | 0.051 |
| Ca | 0.249 | 0.214 | 0.108 | 0.124 |
| $\sigma$Ca | 0.055 | 0.039 | 0.024 | 0.024 |
| Sc | 0.076 | 0.036 | −0.049 | −0.053 |
| $\sigma$Sc | 0.043 | 0.074 | 0.064 | 0.069 |
| Ti | 0.291 | 0.261 | 0.093 | 0.127 |
| $\sigma$Ti | 0.042 | 0.049 | 0.037 | 0.039 |
| V | 0.138 | 0.153 | 0.049 | 0.106 |
| $\sigma$V | 0.086 | 0.063 | 0.036 | 0.035 |
| Cr | 0.006 | −0.067 | −0.022 | −0.005 |
| $\sigma$Cr | 0.098 | 0.053 | 0.092 | 0.139 |
| Mn | −0.402 | −0.429 | −0.405 | −0.414 |
| $\sigma$Mn | 0.052 | 0.036 | 0.027 | 0.021 |
| Co | 0.025 | 0.020 | −0.046 | −0.064 |
| $\sigma$Co | 0.052 | 0.051 | 0.035 | 0.059 |
| Ni | −0.061 | −0.078 | −0.123 | −0.141 |
| $\sigma$Ni | 0.028 | 0.034 | 0.018 | 0.031 |
| Cu | −0.343 | −0.500 | −0.597 | −0.498 |
| $\sigma$Cu | 0.071 | 0.033 | 0.065 | 0.033 |
| Zn | 0.076 | −0.006 | −0.001 | −0.017 |
| $\sigma$Zn | 0.038 | 0.024 | 0.030 | 0.027 |
| Sr | −0.377 | −0.238 | −0.620 | −0.530 |
| $\sigma$Sr | 0.081 | 0.048 | 0.034 | 0.034 |
| Y | 0.058 | −0.024 | −0.237 | −0.180 |
| $\sigma$Y | 0.046 | 0.076 | 0.033 | 0.077 |
| Ba | −0.028 | 0.075 | −0.249 | −0.230 |
| $\sigma$Ba | 0.048 | 0.048 | 0.028 | 0.030 |
| La | ≤0.308 | ≤0.017 | ≤0.111 | ≤−0.035 |
| Ce | ≤0.060 | ≤0.250 | ≤0.055 | ≤0.023 |
| Nd | ≤0.242 | ≤0.262 | ≤−0.166 | ≤−0.064 |
| Sm | ≤0.741 | ≤0.587 | ≤0.364 | ≤0.459 |

Cr, Mn, Fe, Co, Ni, Cu, Zn, Sr, Y, Ba, La, Ce, Nd, and Sm), were estimated. We have used the differential results, along with the absolute measurements of the standard star, and estimated [X/Fe] ratios. The [X/Fe] ratios and their errors are displayed in Table 1; it is important to stress that the abundances of O, La, Ce, Nd, and Sm are upper limits, as they were derived from very weak lines.

We tested the possibility of enhanced rotation in one of the stars by estimating the vsini of stars HD 103095, HD 134439, and HD 134440. We chose four very well-defined Fe I lines (4602, 5250, 4944, and 5247 Å) and performed a spectral synthesis to the lines (with the stellar parameters and Fe abundance previously estimated via equivalent widths), varying the vsini until achieving the best fit. The synthesis was done with the radiative transfer code MOOG (Sneden 1973), using the 'r' option for the line smoothing, which includes the limb darkening (which we adopted to be 0.6), the FWHM of the Gaussian (=0.07 Å, based on the resolution of our spectra), the macroturbulent velocity (the average result from the following equations: $V_{\rm macro} = 13.499 - 0.00707 \times T_{\rm eff} + 9.2422 \times 10^{-7} \times T_{\rm eff}^2$, $V_{\rm macro} = 3.5 + (T_{\rm eff} - 5777)/650$ and $V_{\rm macro} = 3.5 + (T_{\rm eff} - 5777)/388$; Meléndez et al. 2012), and the vsini.

**Table 2.** Differential abundances [X/H] for HD 134440 minus HD 134439.

| Element | [X/H] | $\sigma$[X/H] |
|---|---|---|
| C | 0.090 | 0.043 |
| Na | 0.116 | 0.064 |
| Mg | 0.008 | 0.019 |
| Si | 0.122 | 0.054 |
| K | 0.042 | 0.046 |
| Ca | 0.065 | 0.024 |
| Sc | 0.041 | 0.032 |
| Ti | 0.083 | 0.034 |
| V | 0.106 | 0.035 |
| Cr | 0.066 | 0.031 |
| Mn | 0.040 | 0.034 |
| Co | 0.028 | 0.043 |
| Ni | 0.031 | 0.024 |
| Cu | 0.148 | 0.040 |
| Zn | 0.032 | 0.031 |
| Sr | 0.139 | 0.034 |
| Y | 0.106 | 0.066 |
| Ba | 0.068 | 0.027 |

Subsequently, we performed a new differential analysis on star HD 134440, with star HD 134439 as the standard star. In this analysis, we used the model stellar parameters previously determined (Table 1). So the abundance differences were estimated through the comparison of each measured line of HD 134440 against the same line on HD 134439. The abundance differences we estimated in this new comparison can be seen in Table 2. This new analysis is used only to directly compare the abundances of the binary pair and our interpretation on planet signatures (Section 4.4).

## 4 DISCUSSION

### 4.1 Comparison with the low- and high-$\alpha$ NS10/11 populations.

In Figs 1 and 2, we show the [X/Fe] ratios of O, Na, Mg, Si, Ca, Ti, Cr, Ni, Mn Cu, Zn, Y, Ba, Sc, La, Ce, and Nd, along with abundances of the low- and high-$\alpha$ populations from NS10 and NS11, complemented with abundances from Fishlock et al. (2017) and Ramírez et al. (2012) (we refer to the above four references as 'NS'), and a sample of abundances from stars of dSphs, from the works of Shetrone et al. (2003), Geisler et al. (2005), Monaco et al. (2005), and Letarte et al. (2010).

We also show both ours and NS measurements of the star HD 163810. Although the measurement from the different studies are similar, there is a small offset between the NS measurements and ours. We averaged this difference for the $\alpha$-elements [including oxygen, which uses the same NLTE corrections as the work of Ramírez et al. (2012)] and found a mean offset of 0.05 dex with a standard deviation of 0.04 dex.

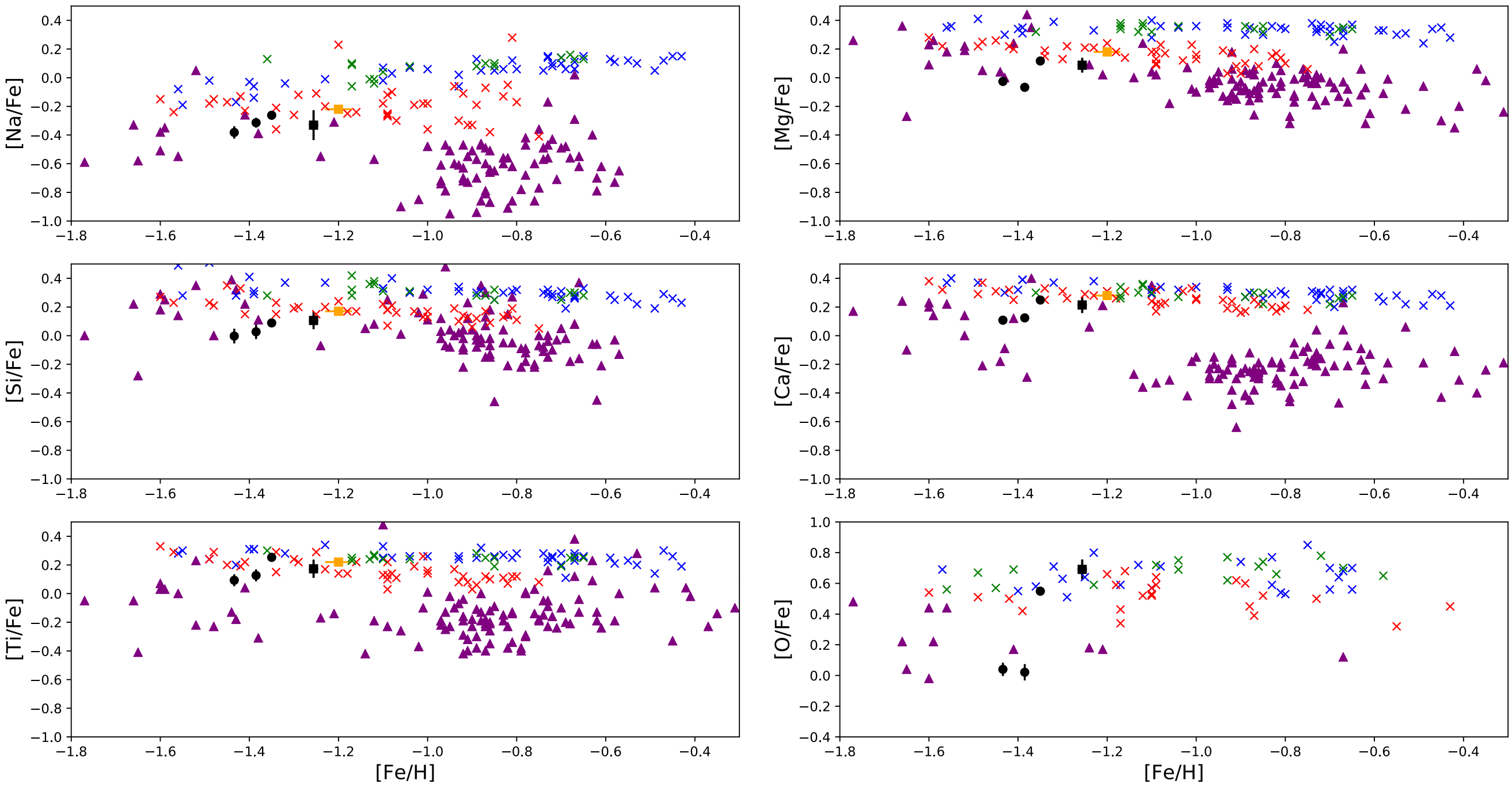

**Figure 1.** [X/Fe] ratios for $\alpha$-elements, Ti and Na. Stars HD 103095, HD 134439, and HD 134440 are plotted as black circles. The standard HD 103095 is the most metal rich of the three stars. The squares represent the star HD 163810 (black and orange colours refer to our measurements and those from NS, respectively). The blue, red, and green crosses are the high-$\alpha$, low-$\alpha$, and thick disc populations of NS. The purple triangles are dSphs (Carina, Sculptor, Fornax, and Sagittarius) stars from Shetrone et al. (2003), Geisler et al. (2005), Monaco et al. (2005), and Letarte et al. (2010).

The black circles are the abundance pattern of stars HD 103095, HD 134439, and HD 134440. The abundance pattern of star HD 103095 closely follows that of the low-$\alpha$ population from NS10. There is a remarkable difference between this star and the pair HD 134439 and HD 134440, specially the [O/Fe] ratio. Our very low oxygen abundance for this pair also agrees with the previous analysis by Chen et al. (2014), who used OH lines in the ultraviolet, in which the HD 134439/HD 134440 binary have [O/Fe] = $-0.26 \pm 0.13$, which is below the most deficient $\alpha$-element in their analysis.

The abundance patterns of HD 134439 and HD 134440 are not compatible with the low-$\alpha$ population of NS for O, Mg, Si, Ca, as can be seen in Fig. 1. Even after adding the systematic difference of 0.05 dex between our measurements and the values from NS, this metal-poor pair still shows [X/Fe] ratios below those of the $\alpha$-poor population of NS10. The striking very low-$\alpha$ abundance pattern that we have determined has been previously reported by Chen et al. (2014, table 12) for O, Mg, Si and Ca, but while they compared the abundance pattern of the pair HD 134439/HD 134440 to the overall abundance pattern of halo stars, we have made a more precise comparison, using a line-by-line analysis relative to the star HD 103095, which has similar stellar parameters, and with observations obtained with the same observational set-up. Also, our comparison of the $\alpha$-poor star HD 163810 from NS indicates that HD 103095 is part of the low-$\alpha$ population. Thus, our work reinforces the previous results by Chen et al. (2014) and demonstrates that the binary HD 134439/HD 134440 is definitely below the low-$\alpha$ population from NS10.

It is important to stress that, although the behaviour of [Ti/Fe] can be similar to the [$\alpha$/Fe] ratios, the nucleosynthetic origin of Ti is still a matter of debate (e.g. Nomoto et al. 2006; The et al. 2006; Clayton 2007; Wongwathanarat et al. 2017). For Na, Ti, Ni, and Zn, the abundances are consistent with the low-$\alpha$ population, with their ratios being slightly below the lower envelope defined by the low-$\alpha$ stars. Both members of the binary system do not show any sizable difference in the abundance pattern of other elements, such as Cr, Mn, and heavy elements Y, Ba, La, and Ce, when compared to the low- and high-$\alpha$ populations.

Albeit the low O and Mg may be suggestive of an origin in a halo globular cluster, the low Na in the pair discards such an hypothesis, as discussed by Chen et al. (2014).

### 4.2 Comparison with dwarf spheroidal galaxies

Studies of chemical abundances in dSph (e.g. Sbordone et al. 2007; Frebel et al. 2016; Ji et al. 2016b; Suda et al. 2017) show that the stars born in these environments have a distinctive abundance pattern of low [$\alpha$/Fe] in a broad range of metallicities, from almost solar to the ultra metal-poor region. Other elements such as Sc also show low abundance ratios.

The abundance pattern of different dSph can be used to better constrain the birth place of the HD 134439/HD 134440 pair. A first approach is to study the position of the knee in different extragalactic systems. The 'knee' is a drop in the $\alpha$-element content due to the different time-scales between type Ia and type II supernovae (Tinsley 1979; Mateucci & Brocato 1990). The particular time-scales of different environments influences the position of the knee in the [$\alpha$/Fe] versus [Fe/H] diagram, which can be used as a chemical tag for the formation environment of the stars. Suda et al. (2017) studied the location of the knee in the Milky Way, Fornax, Sculptor, and Draco, showing that the knees for the above dSphs are at lower metallicities than that of the Milky Way (see table 2 of Suda et al. 2017). In this context, the distinctive $\alpha$-element pattern of HD 134439/HD 134440 resembles the pattern of a dSph with a [$\alpha$/Fe] knee below [Fe/H] $\leq -1.5$, as previously suggested by Chen & Zhao (2006) and Chen et al. (2014).

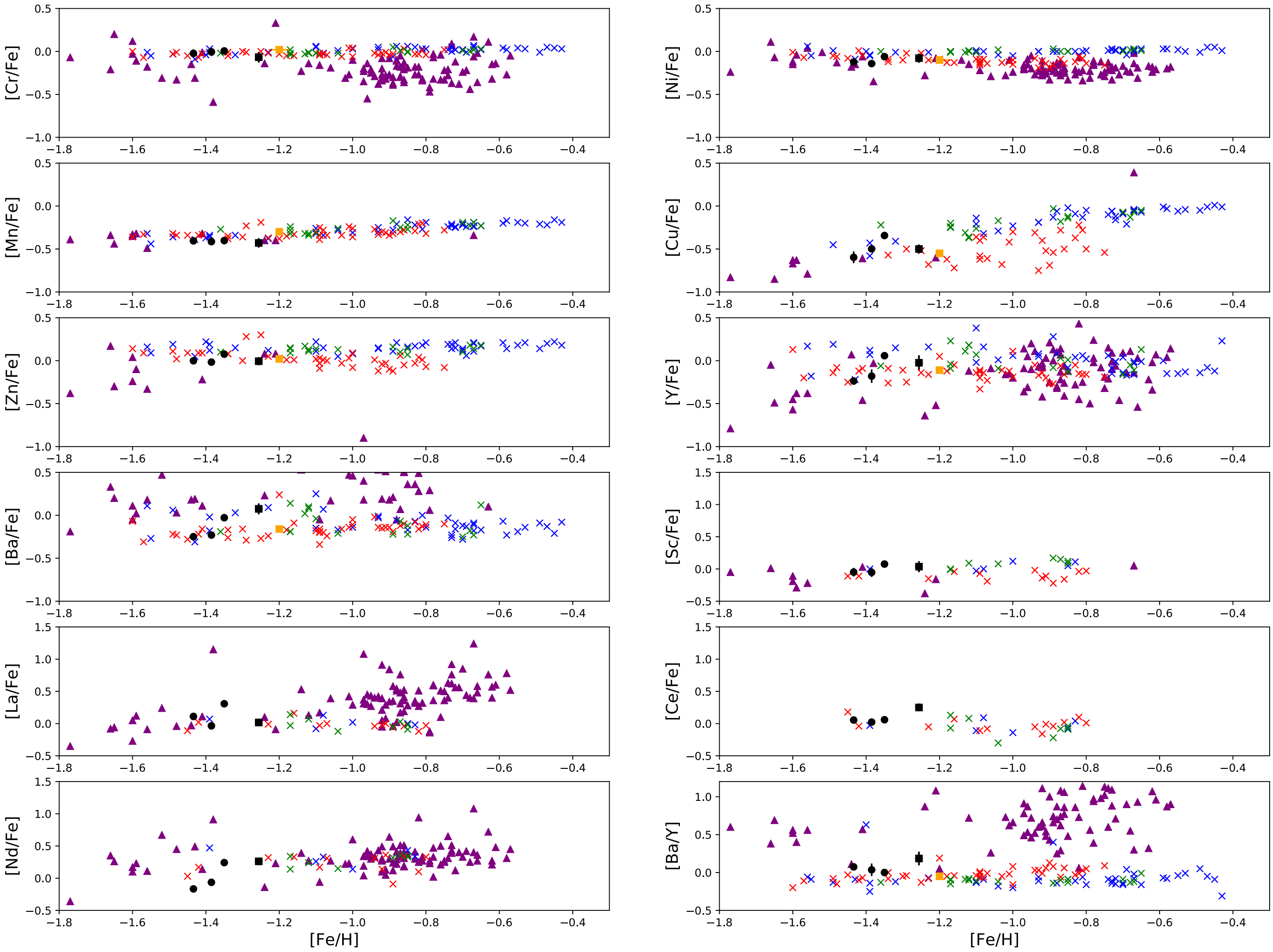

**Figure 2.** [X/Fe] ratios as in Fig. 1 for Cr, Mn, Ni, Cu, Zn, Y, Ba, Sc, La, Ce, Nd, and [Ba/Y].

Spitoni et al. (2016) argues that the best chemical element to tag the birth environment of a star might be barium. From their models, there is a distinctive difference in barium from dSphs, ultra-faint dwarf galaxies (UfDs), and the Milky Way halo. From our measurements, however, the [Ba/Fe] ratio for the binary pair is consistent with the low-$\alpha$ stars (which are believed to be accreted stars) but not consistent with the several measurements of dSphs. Barium has more than one nucleosynthetic origin (it can be produced both in s and r-processes) and the barium excess or lack of excess might not exclusively indicate the formation environment; it may be more important to distinguish the main sources of nucleosynthesis in a given environment.

The [Ba/Y] ratio is also a good indicator of the origin of a given star. A high enough value can be linked to a high SN II environment, with not enough time for the rise of s-process based on AGB stars, like is commonly associated with dSphs (Tolstoy et al. 2009; Chen et al. 2014). This can be clearly seen in our Fig. 2, in which the [Ba/Y] ratio of dSphs stars are clearly super-solar, reaching 1 dex for the higher metallicity regime. Although our results are a bit higher than reported by Chen et al. (2014) and Chen & Zhao (2006) ([Ba/Y] = −0.01, −0.05 for HD 134439 and HD 134440, respectively, against [Ba/Y] = −0.13, −0.08 found by Chen et al. (2014), for HD 134439 and HD 134440, respectively), our result is consistent with theirs as the ratios we report are sub-solar.

As can be seen in the last panel of Fig. 2, the ratios we found are somewhat above the NS11 measurements (although within the errors), as is also the case of our measurements of the comparison star HD 163810. If we subtract from our other objects the offset between our measurements and that by NS11 for HD 163810, the [Ba/Y] ratios for HD 103095, HD 134439, and HD 134440 agree perfectly with the NS11 ratios, thus being in agreement with the low-$\alpha$ and high-$\alpha$ Galactic halo. However, as already pointed by Chen et al. (2014), there is a remarkable offset between these stars and the super-solar ([Ba/Y] $\sim$ 0.6 dex) ratios observed in dSph galaxies, pointing to a birth environment that does not favour a high production of heavy relative to light $n$-capture elements.

We can further constrain the birth environment of these stars using the diagram on fig. 14 of Suda et al. (2017). We adopt the [Eu/Fe] $\leq$ 0.31 for HD 134439 and [Eu/Fe] $\leq$ 0.42 for HD 134440 from Chen et al. (2014) and our measurements for [Ba/Fe], which result in both stars being located on the r-process dominant region, where [Eu/Ba] $\geq$ 0.5.

We used chemical tagging to constrain the birth environment of these stars. We searched for a dSph galaxy with a knee below [Fe/H] $\leq$ −1.5, low [Ba/Y] ratio (indicating low production of heavy relative to low $n$-capture elements) and with stars presenting r-process dominated patterns. We propose that this pair of stars might be from an environment similar to Fornax, as all the above chemical requirements, as well as kinematic constraints, are met.

### 4.3 Trend with condensation temperature

Chen & Zhao (2006) analysed the elemental abundances versus condensation temperature ($T_{\mathrm{cond}}$) of these stars and attributed their abundance pattern to a formation in an environment of low type II supernovae with a high dust-to-gas ratio. On their subsequent study (Chen et al. 2014), they argue against that possibility as the beryllium abundances do not support their previous claim. To verify these results, we show in the upper panel of Fig. 3 a plot of the [X/Fe] abundances versus condensation temperature for both stars. As the slopes are not significant ($3.55 \times 10^{-4} \pm 2.42 \times 10^{-4}$ and $-5.51 \times 10^{-6} \pm 2.49 \times 10^{-4}$, for the blue and black fits in the upper panel), no clear trend with condensation temperature can be seen from this data. The large scatter is likely caused by the comparison of halo stars (with a distinct abundance pattern nonetheless) to the Sun.

Instead of using the Sun as the standard, a more appropriate comparison can be done using a star with a closer nucleosynthetic history, as is the case of HD 103095. Thus, we present the differential abundances HD 134439/HD 134440 − HD 103095. In that case the scatter is considerably reduced and no correlation with condensation temperature (middle and lower panels of Fig. 3) is observed (slopes of $-1.71 \times 10^{-4} \pm 1.00 \times 10^{-4}$ and $-4.54 \times 10^{-5} \pm 8.87 \times 10^{-5}$, significant only at the $\sim 1.7\sigma$ and $\sim 0.5\sigma$ levels for HD 134439 and HD 134440, respectively). Thus, the birth environment of the pair do not have a significant dust effect, as also concluded by Chen et al. (2014) through the analysis of the Be abundance, unlike previous suggestions by Chen & Zhao (2006), who compared the binary pair to the Sun. Our comparison to a standard metal-poor halo star of similar metallicity is more appropriate than the comparison to the Sun ([X/Fe]), which is a thin disc star with quite a different nucleosynthetic history.

### 4.4 Planet signature

It has been claimed that the low-$\alpha$ abundance along with a trend in condensation temperature can be indicative of the engulfment of planetesimals. Chen et al. (2014) argued against this possibility as it is very unlikely that both stars of the pair engulfed the same mass of planetesimals as they have very similar abundance patterns. However, as can be seen in Fig. 4, there is a mean difference of $0.06 \pm 0.01$ dex (weighted abundance and weighted error, from C to Ba, and excluding the upper limits) between the abundances of the pair. As these stars were likely born from the same cloud, as also indicated by their kinematics, the abundance difference between these stars should be zero rather than 0.06 dex. Notice that the differences plotted in Fig. 2 were calculated through a new differential analysis of these stars. In order to better constrain the differences between the stars, we used the stellar parameters previously obtained and recalculated the differential abundances between them, using HD 134439 as the standard star; the differential abundances are shown in Table 2.

To test the probability of finding such abundance differences from zero, we performed a $t$-test, to estimate the probability that two different samples have the same averaged value. The control sample for our test is a sample of zeros (assuming that the binary components have the same abundances). We obtained a $T = 7.51$, which is the calculated difference represented in units of standard error. The higher $T$ represents a higher probability of a real difference between the samples. From the test, we obtained a probability $p = 1.03 \times 10^{-8}$ of obtaining such a high $T$ value. Thus, it is extremely unlikely that the abundance difference of these stars is 0.

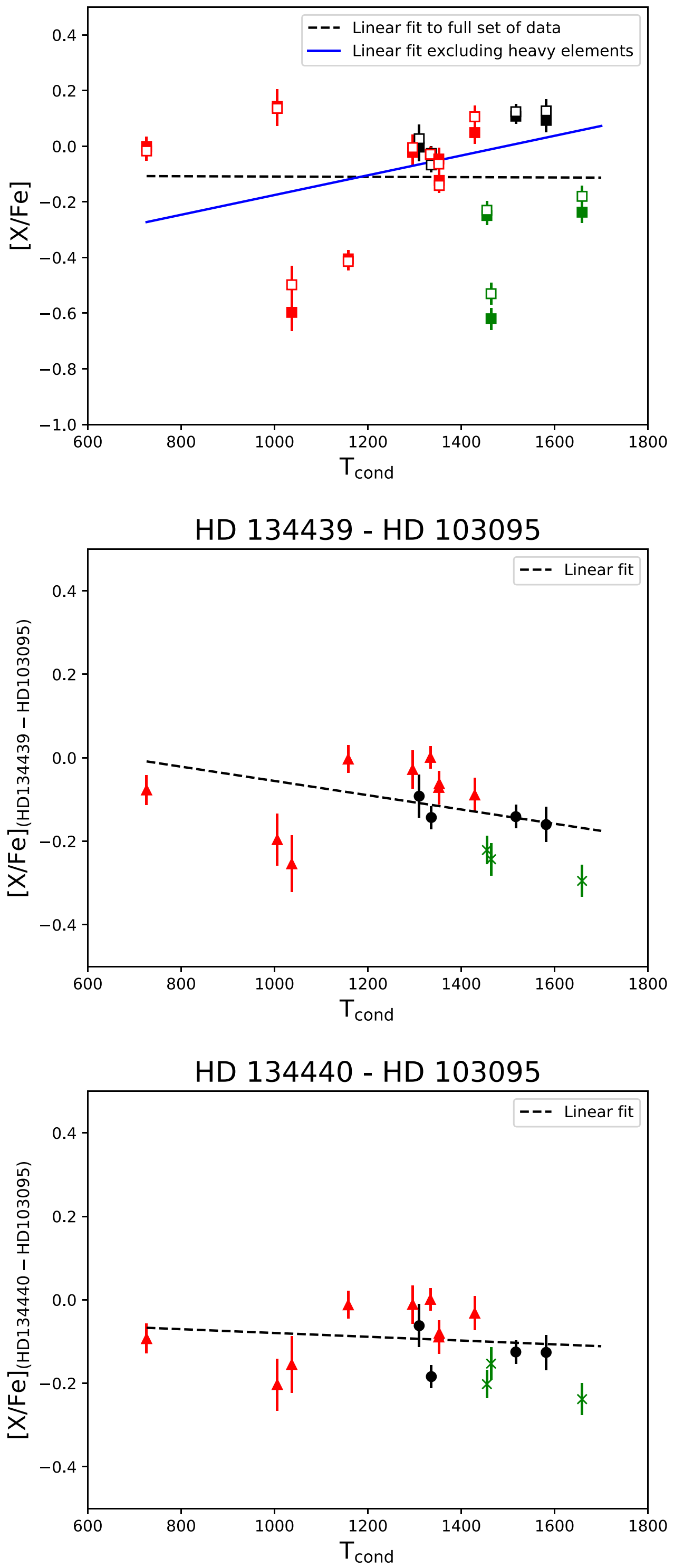

**Figure 3.** Chemical abundances versus $T_{\mathrm{cond}}$. In the top panel, the [X/Fe] for HD 134439 and HD 134440 are shown with filled and open squares, respectively. The middle and lower panels are the differential [X/Fe](HD 134439/HD 134440 − HD 103095) ratios versus $T_{\mathrm{cond}}$. On three panels, the black colours represent the $\alpha$-elements, the red symbols are the iron-peak elements, and the green points represent the $n$-capture elements (Sr, Y, and Ba). The dashed black lines on all panels are the fits to the entire set of abundances and the blue dashed line on the top panel is the fit to the abundances excluding the neutron capture elements.

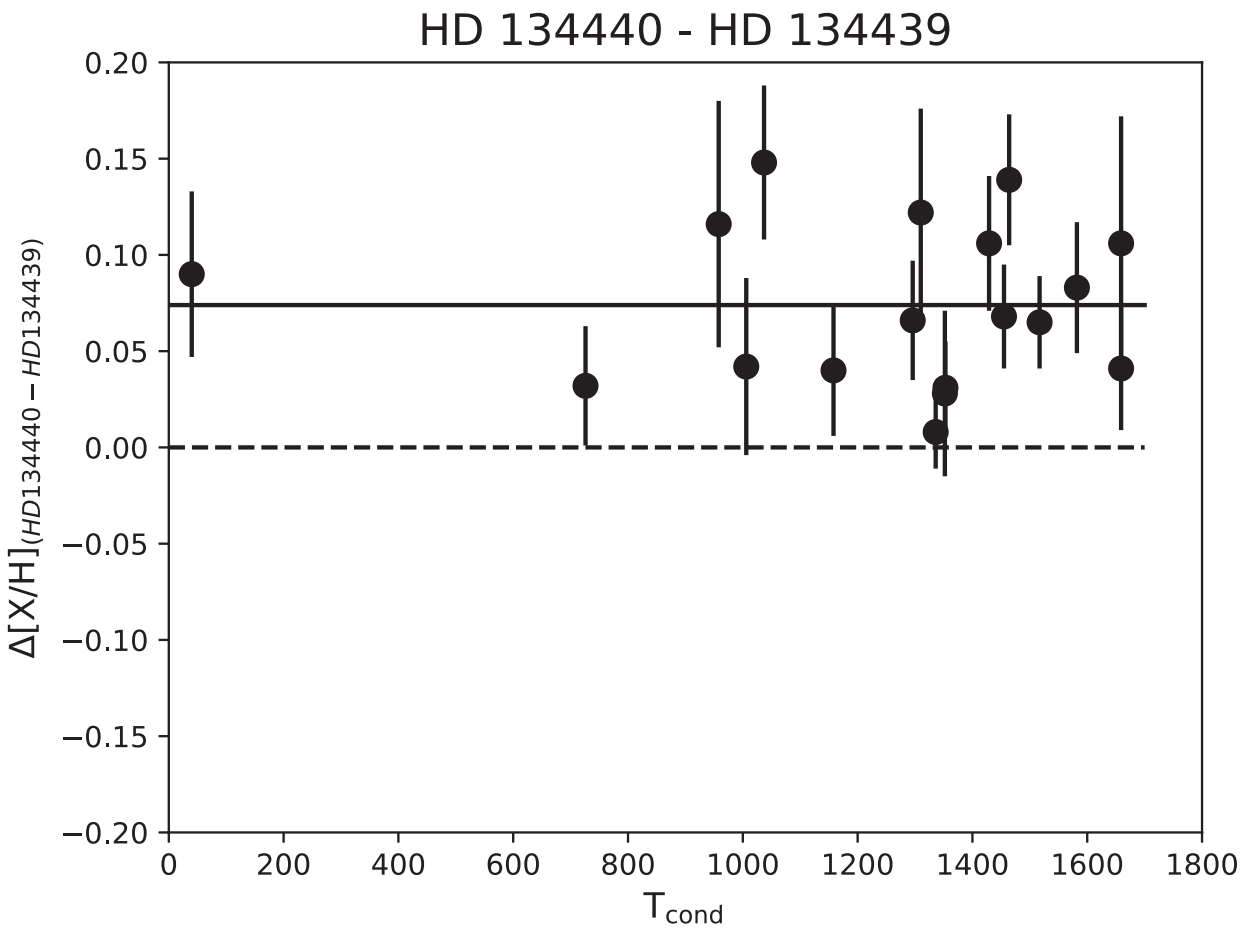

**Figure 4.** Differential Δ[X/H] abundances between the binaries (HD 134440 − HD 134439) from carbon to barium. The solid line is the fit to the entire set of abundances. The dashed line represents the mean difference in chemical composition that the stars would have if they were chemically identical.

We also performed the *t*-test varying 100 000 times the abundances with randomly generated errors from 0 to 0.03 dex, and obtained the average $T = 5.88$ and $p = 2.05 \times 10^{-6}$. Thus, even considering the errors, it is very unlikely that the difference we found could arise just by chance.

Abundance differences have also been found by Chen et al. (2014; table 12), but they found some chemical elements to be less abundant in HD 134440, while others are less abundant in HD 134439; this could be due to differences in the temperature scales between the two works. Chen et al. (2014) adopted the stellar parameters from King (1997), which have a temperature contrast of 215 K between the components, while we found a difference of 138 K. As mentioned by Chen et al. (2014), the newer IFRM temperatures of Casagrande et al. (2010) have a smaller temperature difference between the components, of only 127 K, or about 90 K cooler than the adopted by Chen & Zhao (2006). Notice that our spectroscopic temperature difference (138 K) is only 11 K away from the photometric temperature difference obtained from the IRFM by Casagrande et al. (2010), reinforcing thus the reliability of our temperature scale and our spectroscopic measurements. Also, the fact that all our abundance differences (HD 134440 − HD 134439) are consistently of the same sign, reinforces the reliability of the offset we found between the binaries.

In their study of the binary pair of solar twins 16 Cyg A and B (planet host), Ramírez et al. (2011) also found a similar abundance difference between the components of 0.04 dex. Schuler et al. (2011) did not find abundance differences in 16 Cyg, but further work with higher resolving power corroborated a systematic abundance difference between 16 Cyg A and B (Tucci Maia, Meléndez & Ramírez 2014), a finding also confirmed at lower precision by Mishenina et al. (2016) and at extremely high precision by two recent works using spectra with $R = 115\,000$ (Nissen et al. 2017) and $R = 160\,000$ (Tucci Maia et al. 2018). As both components were formed from the same natal cloud, and both are twins, the abundance differences are likely related to planets. Other precise works on twin binaries with planets show abundance differences between the components of the binary pair, which are likely related to planets (Biazzo et al. 2015; Ramírez et al. 2015; Teske et al. 2016a; Teske, Khanal & Ramírez 2016b; Saffe et al. 2017).

Sun-like stars deplete lithium as they age (e.g. Do Nascimento et al. 2009; Denissenkov 2010; Carlos, Nissen & Meléndez 2016; Reddy & Lambert 2017), but planets could also have an effect on the stellar Li abundances, causing either an enhancement (e.g. Sandquist et al. 2002; Aguilera-Gómez et al. 2016; Carlos et al. 2016; Meléndez et al. 2017; Saffe et al. 2017) or a depletion (e.g. Théado & Vauclair 2012; Deal, Richard & Vauclair 2015; Gonzalez 2015). Although planet accretion should increase the lithium content of the star, depending on the conditions of the accretion it could actually deplete Li due to thermohaline mixing (Théado & Vauclair 2012) or to increased rotation due to transfer of angular momentum.

The HD 134439/HD 134440 lithium abundances were studied by King (1997), who found a lithium difference of 0.6 dex between them (HD 134440 being Li depleted), and claimed that the observed Li difference could be due to large differences in temperature (implying a significant mass difference) between the two components. We have also synthesized the lithium 6707.7 Å line and found similar results; for HD 103095 $\epsilon(\mathrm{Li}) = +0.37 \pm 0.15$, on HD 1334439, $\epsilon(\mathrm{Li}) = +0.55 \pm 0.15$, and $\epsilon(\mathrm{Li}) \leq -0.05$ in HD 134440. However, we find no significant difference in the stellar parameters to account for the lithium differences. Using the q2 code (Ramírez et al. 2014), we estimated the masses and ages of these stars and both are very similar: 0.59 and 0.58 $M_\odot$ and 9.9 and 9.4 Gyr for HD 134439 and HD 134440, respectively.

Thus, we see no evidence that would corroborate a stellar evolution explanation for the low lithium content on HD 134440. Instead, the low Li in HD 134440 could be due to a planet engulfment event. As mentioned above, under some conditions a planet engulfment event could trigger thermohaline convection and cause Li destruction (Deal et al. 2015). Interestingly, Chen et al. (2014) barely detected Be in HD 134439, while it was not present at all in HD 134440. This means that the induced extra-mixing due to the possible planet engulfment event was high enough that not only Li but also Be was depleted in HD 134440. The planet engulfment would also lead to an increase in the abundance of all metals, as we seem to detect in HD 134440, which seem enhanced in all chemical elements, in comparison to its companion HD 134439.

It is unclear whether the rotation rate would be largely enhanced or not due to a planet engulfment event. In order to verify a possible increased rotation rate in HD 134440, we estimated vsini for HD 103095, HD 134439, and HD 134440, as previously mentioned in Section 3. We found vsini = 1.2 ± 0.5, 1.9 ± 0.5, and 2.0 ± 0.5 km $\mathrm{s}^{-1}$ for HD 103095, HD 134439, and HD 134440, respectively. Thus, we see no significant difference in the rotation velocities of the binary pair, which means that the planet engulfment did not affect much the rotation rate of HD 134440. However, we cannot fully discard an increase in the rotation rate due to the unknown inclination angles.

Johnson & Li (2012) estimated the critical metallicity for planet formation as $[\mathrm{Fe/H}]_{\mathrm{crit}} = -1.5 + \log(r/1\,\mathrm{au})$, where $r$ is the distance of the planet to its host star in astronomical units (au). As the metallicity of the binary pair is [Fe/H] = −1.4, it means that it could form planets inside ∼1.3 au, albeit this distance depends on the conditions for planet formation at low metallicities, which are still uncertain.

In light of the aforementioned discussion, we seem to detect, for the first time, the effect of planet engulfment in a metal-poor star. The star HD 134440 seems richer in the measured chemical elements, as the result of the planet engulfment, which may also explain the Li and Be difference amongst the two components of this binary pair.

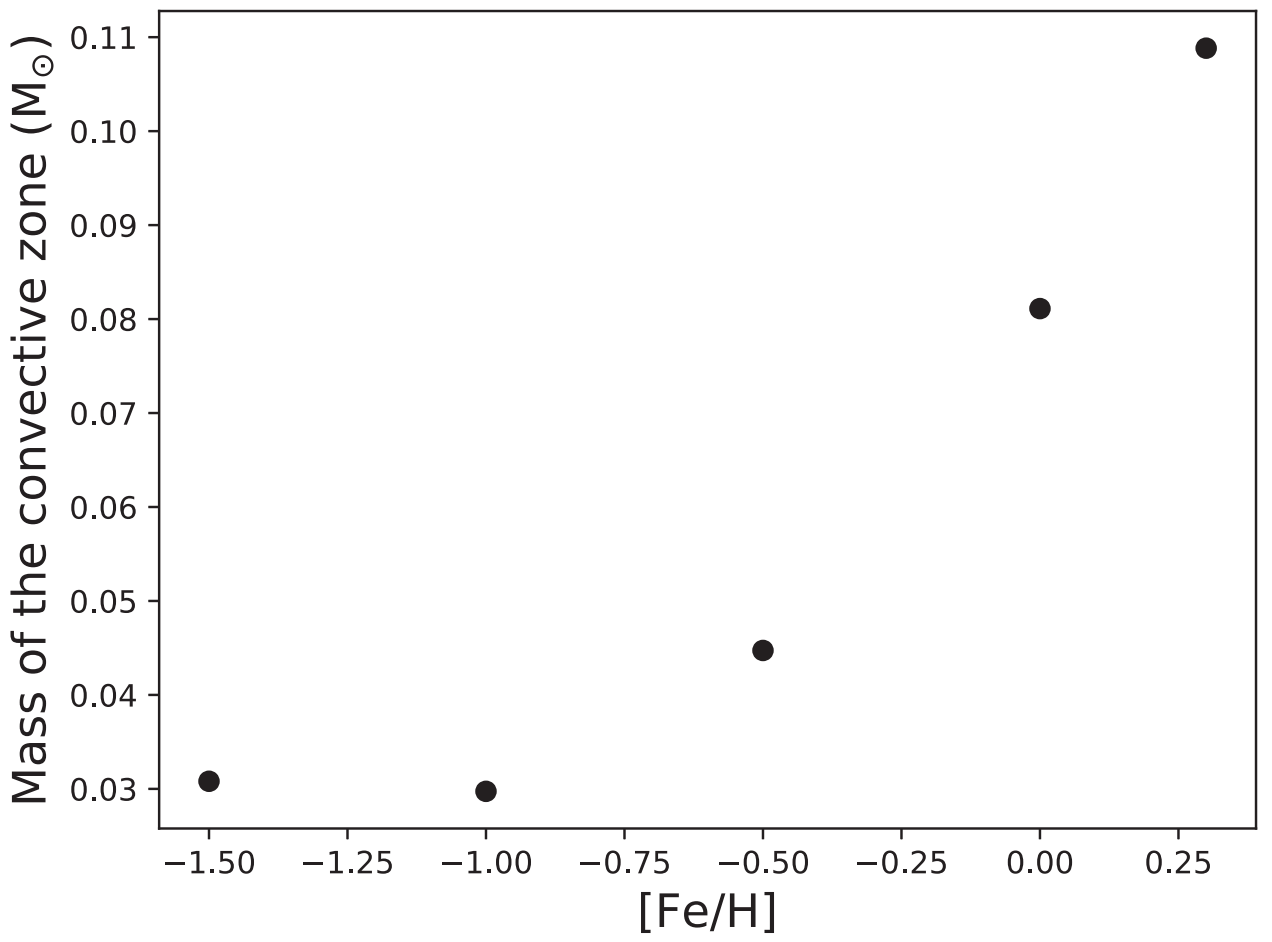

**Figure 5.** Mass of the convective zone for stars of 0.6 $M_\odot$ and 9.5 Gyr, at different metallicities, from the YAPSI tracks with a mixing length of 1.92.

As the two stars (HD 134439 and HD 134440) are most likely from the same birth cloud, we assume that the difference in chemical abundance is due to a planet that might have formed and later engulfed. We estimate a possible mass of the engulfed planet using the same toy model proposed by Ramírez et al. (2011, equation 1). We assume the metal composition of the planet (Z/X) to be five times the initial composition of the star, which we adopt to be the same metallicity we observe today in HD 134439. With those assumptions and using a mass of 0.03 $M_\odot$ for the convective zone of HD 134440 (from the YAPSI[4] evolutionary tracks, see Fig. 5), we estimate a planetary mass of $M \sim 0.9\,M_J$ that could have been engulfed by HD 134440. Notice that as the convection zone is small at low metallicities (Fig. 5), planets can have an important effect on the chemical composition of metal-poor stars.

## 5 CONCLUSIONS

We conclude that the binary pair HD 134439/HD 134440 do not belong to either the low-$\alpha$ nor the high-$\alpha$ halo populations proposed by NS10, but follow more closely the $\alpha$ and the $r$-process dominant pattern of stars seen in dwarf galaxies such as Fornax. We also see evidence for the possible accretion of a 0.9 $M_J$ planet by HD 134440. If confirmed, it could corroborate that planetary formation can occur down to metallicities of [Fe/H] $\sim -1.5$.

## ACKNOWLEDGEMENTS

H.R. thanks a CAPES fellowship and the CAPES PDSE programme (88881.132145/2016-01). J.M. thanks support by FAPESP (2012/24392-2 and 2014/18100-4) and CNPq (Productivity fellowship). We also thank Dr. Akito Tajitsu for his important support and initial data reduction during the observations with the Subaru Telescope. This study is based (in part) on data collected at Subaru Telescope, which is operated by the National Astronomical Observatory of Japan.

[4] http://www.astro.yale.edu/yapsi/download_grids.html

## REFERENCES

Aguilera-Gómez C., Chanamé J., Pinsonneault M. H., Carlberg J. K., 2016, ApJ, 829, 127
Aoki W. et al., 2009, A&A, 502, 569
Biazzo K. et al., 2015, A&A, 583, A135
Bonifacio P., Hill V., Molaro P., Pasquini L., DiMarcantonio P., Santin P., 2000, A&A, 359, 663
Bonifacio P., Sbordone L., Marconi G., Pasquini L., Hill V., 2004, A&A, 414, 503
Carlos M., Nissen P. E., Meléndez J., 2016, A&A, 587, A100
Carney B. W., Laird J. B., Latham D. W., Aguilar L. A., 1996, AJ, 112, 668C
Casagrande L., Ramírez I., Meléndez J., Bessell M., Asplund M., 2010, A&A, 512, A54
Chen Y. Q., Zhao G., 2006, MNRAS, 370, 2091
Chen Y., King J. R., Boesgaard A. M., 2014, PASP, 126, 1010C
Clayton D., 2007, Handbook of Isotopes in the Cosmos. Cambridge Univ. Press, Cambridge
Cohen J. G., Huang W., 2009, ApJ, 701, 1053
Deal M., Richard O., Vauclair S., 2015, A&A, 584, A105
Denissenkov P. A., 2010, ApJ, 719, 28
Do Nascimento J. D., Jr, Castro M., Meléndez J., Bazot M., Théado S., Porto de Mello G. F., de Medeiros J. R., 2009, A&A, 501, 687
Fabrizio M. et al., 2015, A&A, 580, A18
Fishlock C. K., Yong D., Karakas A. I., Alves-Brito A., Meléndez J., Nissen P. E., Kobayashi C., Casey A. R., 2017, MNRAS, 466, 4672
Frebel A., Norris J. E., Gilmore G., Wyse R. F. G., 2016, ApJ, 826, 110
Geisler D., Smith V. V., Wallerstein G., Gonzalez G., Charbonnel C., 2005, AJ, 129, 1428
Gonzalez G., 2015, MNRAS, 446, 1020
Gustafsson B., Edvardsson B., Eriksson K., Jørgensen U. G., Nordlund Å., Plez B., 2008, A&A, 486, 951
Ji A. P., Frebel A., Simon J. D., Chiti A., 2016b, ApJ, 830, 93
Johnson J. L., Li H., 2012, ApJ, 751, 81
King J. R., 1997, AJ, 113, 2302
Lemasle B. et al., 2012, A&A, 538, A100
Lemasle B. et al., 2014, A&A, 572, A88
Letarte B. et al., 2010, A&A, 523, A17
Matteucci F., Brocato E., 1990, ApJ, 365, 539
Meléndez J. et al., 2012, A&A, 543, A29
Meléndez J. et al., 2017, A&A, 597, A34
Mishenina T., Kovtyukh V., Soubiran C., Adibekyan V. Z., 2016, MNRAS, 462, 1563
Monaco L., Bellazzini M., Bonifacio P., Ferraro F. R., Marconi G., Pancino E., Sbordone L., Zaggia S., 2005, A&A, 441, 141
Nissen P. E., Schuster W. J., 2010, A&A, 511, L10 (NS10)
Nissen P. E., Schuster W. J., 2011, A&A, 530, A15 (NS11)
Nissen P. E., Silva Aguirre V., Christensen-Dalsgaard J., Collet R., Grundahl F., Slumstrup D., 2017, A&A, 608, A112
Noguchi K. et al., 2002, PASJ, 54, 855
Nomoto K., Tominaga N., Umeda H., Kobayashi C., Maeda K., 2006, Nucl. Phys. A, 777, 424
Ramírez I., Meléndez J., Cornejo D., Roederer I. U., Fish J. R., 2011, AJ, 740, 76
Ramírez I., Meléndez J., Chanamé J., 2012, ApJ, 757, 164
Ramírez I. et al., 2014, A&A, 572, A48
Ramírez I. et al., 2015, ApJ, 808, 13
Reddy A. B. S., Lambert D. L., 2017, ApJ, 845, 151
Reggiani H., Meléndez J., Yong D., Ramírez I., Asplund M., 2016, A&A, 586, A67
Saffe C., Jofré E., Martioli E., Flores M., Petrucci R., Jaque Arancibia M., 2017, A&A, 604, L4
Sandquist E. L., Dokter J. J., Lin D. N. C., Mardling R. A., 2002, ApJ, 572, 1012
Sbordone L., Bonifacio P., Buonanno R., Marconi G., Monaco L., Zaggia S., 2007, A&A, 465, 815

Schuler S. C., Cunha K., Smith V. V., Ghezzi L., King J. R., Deliyannis C. P., Boesgaard A. M., 2011, ApJ, 737, L32
Shetrone M. D., Bolte M., Stetson P. B., 1998, AJ, 115, 1888
Shetrone M. D., Côté P., Sargent W. L. W., 2001, ApJ, 548, 592
Shetrone M. D., Venn K., Tolstoy A., Primas F., Hill V., Kaufer A., 2003, AJ, 125, 684
Shigeyama T., Tsujimoto T., 2003, ApJ, 598, L47
Sitnova T. et al., 2015, ApJ, 808, 148
Sneden C. A., 1973, PhD thesis, Univ. Texas
Spitoni E., Vincenzo F., Matteucci F., Romano D., 2016, MNRAS, 458, 2541
Suda T. et al., 2017, PASJ, 69, 76
Teske J. K., Shectman S. A., Vogt S. S., Díaz M., Butler R. P., Crane J. D., Thompson I. B., Arriagada P., 2016a, AJ, 152, 167
Teske J. K., Khanal S., Ramírez I., 2016b, ApJ, 819, 19
Théado S., Vauclair S., 2012, ApJ, 744, 123
The L.-S. et al., 2006, A&A, 450, 1037
Tinsley B. M., 1979, ApJ, 229, 1046
Tolstoy E., Hill V., Toii M., 2009, ARA&A, 47, 371
Tomkin J., 1972, MNRAS, 156, 349
Tucci Maia M., Meléndez J., Ramírez I., 2014, AJ, 790, L25
Tucci Maia M., Meléndez J., Spina L., Lorenzo-Oliveira D., 2018, MNRAS, submitted
Venn K. A. et al., 2012, ApJ, 751, 102
Wongwathanarat A., Janka H.-T., Müller E., Pllumbi E., Wanajo S., 2017, ApJ, 842, 13
Yong D. et al., 2013, MNRAS, 434, 3542

## SUPPORTING INFORMATION

Supplementary data are available at *MNRAS* online.

**Table A1.** Example of the linelist used for the abundances determinations.

Please note: Oxford University Press is not responsible for the content or functionality of any supporting materials supplied by the authors. Any queries (other than missing material) should be directed to the corresponding author for the article.

## APPENDIX: LINELIST

**Table A1.** Example of the linelist used for the abundances determinations, formatted to be used with the radiative transfer code MOOG (Sneden 1973), and also include the hyperfine splitting, indicated by the negative wavelengths. The full linelist can be found online.

| Wavelength (Å) | Species | EP (eV) | log(*gf*) (dex) | EW_HD 103095 (mÅ) | EW_HD 134439 (mÅ) | EW_HD 134440 (mÅ) | EW_HD 163810 (mÅ) |
|---|---|---|---|---|---|---|---|
| 4445.471 | 26 | 0.087 | −5.441 | 25.6 | 24.5 | 33.7 | 11.7 |
| 4602.001 | 26 | 1.608 | −3.154 | 50.6 | 47.7 | 56.0 | 32.4 |
| 4779.439 | 26 | 3.415 | −2.020 | 15.3 | 12.3 | 19.5 | 7.7 |
| 4788.757 | 26 | 3.237 | −1.763 | 31.1 | 29.4 | 34.5 | 22.4 |
| 4950.106 | 26 | 3.417 | −1.560 | 35.7 | 33.1 | 39.6 | 25.6 |
| 4994.129 | 26 | 0.915 | −3.080 | 95.9 | 97.9 | 114.2 | 71.0 |
| 5044.211 | 26 | 2.851 | −2.058 | 49.5 | 43.3 | 59.3 | 27.7 |
| 5054.642 | 26 | 3.640 | −1.921 | 16.5 | 13.1 | 19.5 | 9.8 |
| 5127.359 | 26 | 0.915 | −3.307 | 86.4 | 86.1 | 99.6 | 63.9 |
| 5198.711 | 26 | 2.223 | −2.135 | 87.0 | 84.2 | 106.5 | 57.0 |

This paper has been typeset from a TEX/LATEX file prepared by the author.